\documentclass[pra,twocolumn,showpacs,groupedaddress]{revtex4}

\usepackage{dcolumn}
\usepackage{graphicx}
\usepackage{amsmath}
\usepackage{bm}

\usepackage[usenames]{color}

\newcommand{\ud}[1]{{#1^{\dagger}}}
\newcommand{\bra}[1]{\left\langle #1\right|}
\newcommand{\ket}[1]{\left| #1\right\rangle}

\newcommand\Tr{\mathrm{Tr}}
\newcommand{\mean}[1]{\langle#1\rangle}

\begin{document}

\title{Photon correlations from ultra-strong optical nonlinearities}

\author{Alessandro Ridolfo$^{1}$, Elena del Valle$^{1,2}$, and Michael J. Hartmann$^{1,3}$}
\affiliation{$^{1}$Physik Department, Technische Universit\"{a}t M\"{u}nchen, James-Franck-Strasse, 85748 Garching, Germany\\
  $^{2}$F\'{i}sica Te\'{o}rica de la Materia Condensada, Universidad Aut\'{o}noma de Madrid, 28049 Madrid, Spain\\
  $^{3}$Institute of Photonics and Quantum Sciences, Heriot-Watt University, Edinburgh, EH14 4AS, United Kingdom}
\date{\today}

\begin{abstract}
  We study the full field and frequency filtered output photon
  statistics of a resonator in thermal equilibrium with a bath and
  containing an arbitrarily large quartic nonlinearity. According to
  the general theory of photodetection, we derive general input-output
  relations valid for the \emph{ultra-anharmonic} regime, where the
  nonlinearity becomes comparable to the energy of the resonator, and
  show how the emission properties are modified as compared to the
  generally assumed simple \emph{anharmonic} regime. We analyse the
  impact of the nonlinearity on the full statistics of the emission,
  $g^{(2)}$, and its spectral properties. In particular we derive a
  semi-analytical expression for the frequency resolved two-photon
  correlations or two-photon spectrum of the system in terms of the
  master equation coefficients and density matrix. This provides a
  very clear insight into the level structure and emission
  possibilities of the system.
\end{abstract}


\maketitle

\section{Introduction}

The quantum properties of light fields are one of the central objects
studied in Quantum Optics. In this context it has been realized early
on that optical nonlinearities are needed to generate non-classical
output fields from classical input \cite{Walls}. Therefore,
engineering large optical nonlinearities has been a prime goal in
experimental Quantum Optics throughout recent decades. Since the
propagation of light fields in vacuum is described by a linear wave
equation, optical nonlinearities can only appear if light fields
couple to a suitable medium. Hence a strong nonlinearity requires a
strong light matter coupling in the first place. Very recently,
exceptionally strong light matter interactions have been realized in a
variety of solid state optical devices
\cite{guenter09,Niemczyk,Todorov,Schwartz,Hoffman,Scalari}. In fact
these light matter interactions have reached coupling strengths that
are comparable to the energy of the photons that interact with the
matter, leading to a novel regime of light matter coupling that has
been coined the ultra-strong coupling regime.

Ultra-strong light matter couplings in turn will also lead to optical
nonlinearities of unprecedented strength. The characterization of the
physics of optically ultra-nonlinear devices is therefore a very
timely question of high interest
\cite{RidolfoPrl2012,RidolfoPhysScr2013,RidolfoPrl2013,StassiPrl2013}.
Here we investigate the output photon statistics of optical
nonlinearities for the regime where the anharmonicity of their
frequency spectrum becomes comparable to the frequency of single
photons. In our studies we focus on two paradigm examples of optical
nonlinearities, a Kerr nonlinearity \cite{Imamoglu} and a $\chi^{(3)}$
nonlinearity \cite{Scullybook}. For the regime we are interested in, there is a
significant difference between these two examples as no rotating wave
approximation can be applied in the equations of motion
\cite{Drummond80}.

The statistics of output photons for the ultra nonlinear devices we
consider differs dramatically from the physics encountered in standard
regimes where the anharmonicity of the spectrum is small compared to
the photon frequency. There are two main reasons for these marked
differences. First, photon dissipation is strongly modified for ultra
strong nonlinearities. In each dissipation event the system loses a
photon but in contrast to weak nonlinearities the frequency of the
emitted photon strongly depends on how many photons are present inside
the nonlinear device. This frequency dependence of the emission events
needs to be taken into account properly \cite{Blais}. Second, the
frequency dependence of emitted photons also needs to be accounted for
properly in the relation between the field in the device and an output
field impinging on the detector. This requires properly generalized
input-output relations \cite{RidolfoPrl2012}.
 
Ultra-nonlinear devices for microwave photons are for example very
nicely realized in circuit quantum electrodynamics. In particular
Hamiltonians as we consider them are realized in superconducting
qubits, such as transmons \cite{Koch07} or in a transmission line
resonator where the central conductor is intersected by a direct
current superconducting interference device (dc SQUID)
\cite{Leib12,Bourassa12}.

The remainder of the paper is organized as follows.
In section \ref{sec:model} we first introduce the model we consider
which may be analyzed in two versions. In the first version it contains
an utrastrong Kerr nonlinearity for which we discuss the output photon
statistics and spectra in section \ref{sec:kerr}. The photon correlations and spectra
of the second version containing a general $\chi^{(3)}$ nonlinearity that is quartic in the field are
presented in section \ref{sec:full}. In section \ref{sec:attractive} we examine the
photon correlations for the considered model with a negative $\chi^{(3)}$ nonlinearity, since
some typical implementations such as superconducting qubits are described by it.
We then conclude in section \ref{sec:concl}.

\section{Model}
\label{sec:model}

The aim of this work is to study the steady state and emission
statistical properties of a non-linear resonator coupled to a thermal
reservoir. The most general Hamiltonian that takes into account the
nonlinearity of the system, consists of the harmonic part plus a
standard power expansion of the potential energy with coefficients
$U_{2n}$. Here we restrict ourselves to models where the potential is
symmetric around the point where the field vanishes so that only even
powers of the field appear in the expansion of the potential.
Moreover we concentrate most of our discussion on cases where the
nonlinearity is repulsive, i.e. $U_{2n} \ge 0$ for all $n$, as this
ensures that the energy of the system has a lower bound for arbitrary
magnitudes of the $|U_{2n}|$ and thus yields physically meaningful
results for arbitrary field amplitudes.  Setting $\hbar = 1$, this
Hamiltonian reads
\begin{equation} \label{h-full}
  H = \omega_{a} a^{\dagger}a + \sum_{n = 2}^{\infty} U_{2 n} (a + a^{\dagger})^{2 n}\,,
\end{equation}
where $\omega_{a}$ is the bare mode frequency of the resonator and $a$
its annihilation operator.  For moderate field amplitudes, that is,
for regimes with moderate photon numbers, the main physical effects
due to the nonlinearity of the system are well described truncating
such an expression to fourth order, i.e.
\begin{equation} \label{hamiltonian1}
  H_{S} = \omega_{a} a^{\dagger}a + U (a + a^{\dagger})^{4},
\end{equation}
which is the quantized version of the classical Hamiltonian of the
Duffing oscillator \cite{Peano06}.

If the system we analyze is in a
regime of weak perturbation, i.e. the number of total excitations is
small, one argues that the off-diagonal terms in
Eq.~(\ref{hamiltonian1}), like the squeezing terms $(a^{\dagger})^2$
and $a^2$, are negligible as they create (or destroy) more than one
excitation at a time. In this case, a further simplification leads to
the well-known Kerr-nonlinear Hamiltonian,
\begin{equation} \label{hamiltonian2}
  H_{K} = \omega_{a} a^{\dagger}a + U a^{\dagger}a^{\dagger} a a\,.
\end{equation}

The exchange of excitations with a thermal bath with temperature $T$ is
described in a master equation in the Lindblad form,
\begin{equation}\label{master-eq}
  \dot\rho(t) =\mathcal{L} \rho (t) = i [\rho(t), H] + \mathcal{L}_T\rho(t)\,,
\end{equation}
with $\rho (t)$ the density matrix of the resonator and the dot
denoting a time derivative. We will consider $T$ in energy units that
include the Boltzmann constant $k_\mathrm{B}$. The standard expression
used in the literature \cite{breuer} for a thermal bath is
\begin{equation}
  \label{eq:wrong-thermal}
  \mathcal{\tilde  L}_{T}=\gamma_a\Big[(1+\bar n_T)\mathcal{D}_a+\bar n_T\mathcal{D}_{a^\dagger}\Big]
\end{equation}
with $\mathcal{D}_a\rho = \frac{1}{2} (2 a \rho a^{\dagger}-\rho
a^{\dagger} a - a^{\dagger} a\rho)$, $\bar n_T$ the occupation of the
bath at temperature $T$ and $\gamma_a$ the decay rate into the bath at
zero temperature~\cite{GardinerZoller}. However,
Eq.~(\ref{eq:wrong-thermal}) is derived under the assumption that
$U\ll \omega_{a}$, and therefore the steady state it leads to, is
independent of $U$:
\begin{equation}
  \label{eq:FriApr19142616CEST2013}
  \tilde{\rho}_T=\frac{1}{\tilde{Z}}\sum_{n} e^{-\frac{\omega_a n}{T}}\ket{n}\bra{n}\,,
\end{equation}
with $\tilde{Z}=\sum_{n} e^{-\frac{\omega_a n}{T}}$. Importantly,
$\tilde{\rho}_T$ differs from the thermal equilibrium state of an
anharmonic oscillator~\cite{Alicki} with a level structure described
by $H_K$ or $H_S$,
\begin{equation}
  \label{eq:thermaleq}
  \rho_T=\frac{1}{Z} e^{-\frac{H_{\alpha}}{T}}\,,
\end{equation}
for $\alpha = K$ or $\alpha = S$, and $Z = \Tr
[e^{-\frac{H_{\alpha}}{T}}]$. We explore this discrepancy and its
implications in the sequel. In doing so we first focus on the Kerr
Hamiltonian.

\section{Kerr nonlinearity}
\label{sec:kerr}

\begin{figure}
  \centering
  \includegraphics[width=\linewidth]{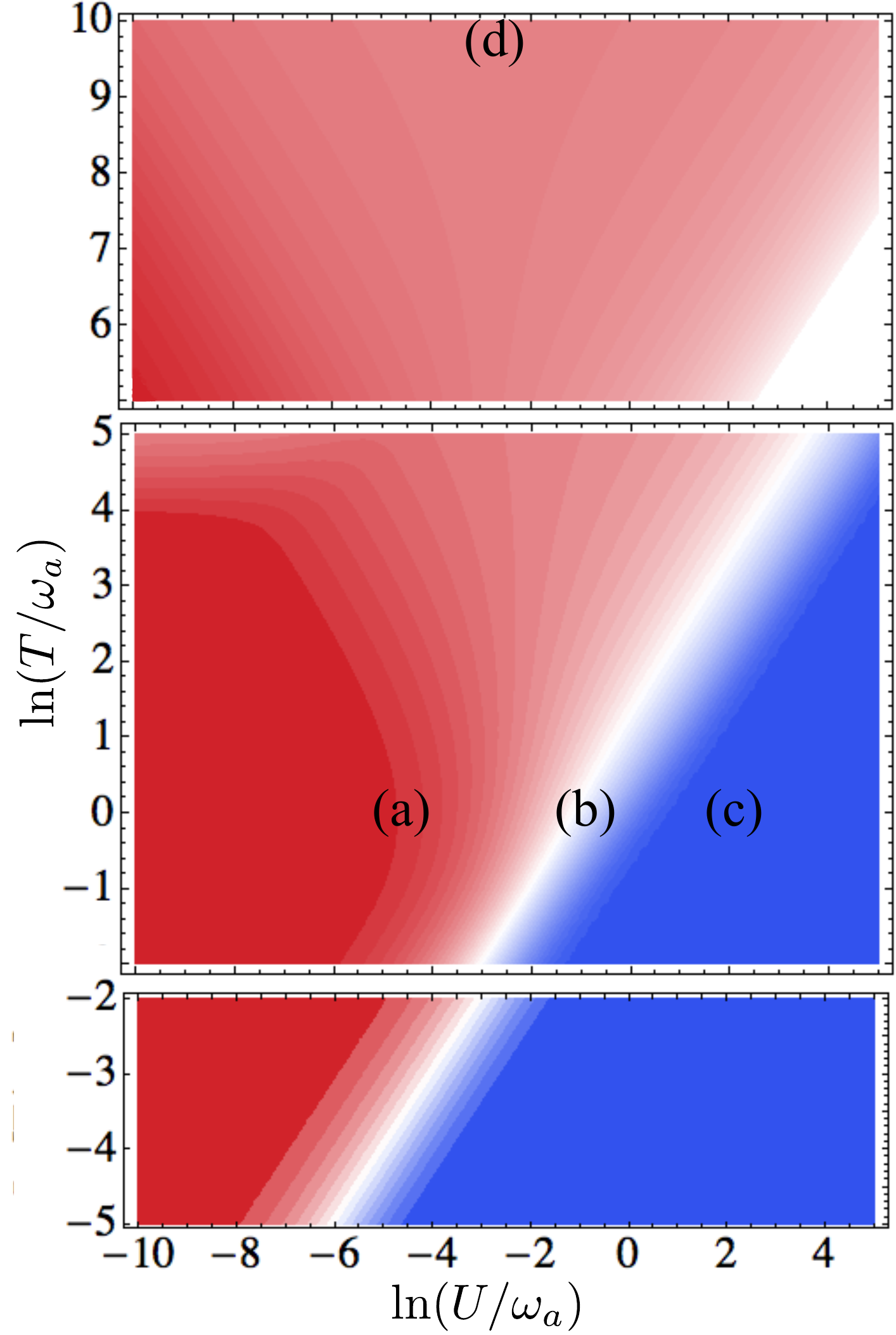}
  \caption{(color online) $g_a^{(2)}$ as a function of temperature and
    nonlinearity for the Kerr Hamiltonian $H_K$. Upper panel: the high
    occupation approximated solution, with the top limit
    $\pi/2$. Middle panel: Numerical solution. Lower panel: the low
    occupation approximated solution. Red corresponds to 2, white to 1
    and blue to 0. The points marked with letters are further
    investigated in Fig. \ref{fig:stats}.}
  \label{fig:kerrg2}
\end{figure}

Since $H_{K}$ is diagonal in the resonator number state basis,
$\langle m|H_{K}|n\rangle=\delta_{mn}\epsilon_{n}$, calculations are
straightforward and even analytical in some limits. For instance,
transition energies between the levels with $\epsilon_{n} = n
\omega_{a} + n(n-1) U$ and $n=0,1,\ldots$ are simply given by $\Delta
\epsilon_{n} =\epsilon_{n}-\epsilon_{n-1}=\omega_{a} + 2 (n-1) U$,
i.e. they increase linearly with $n$ and $U$.

\begin{figure}
  \centering
  \includegraphics[width=\linewidth]{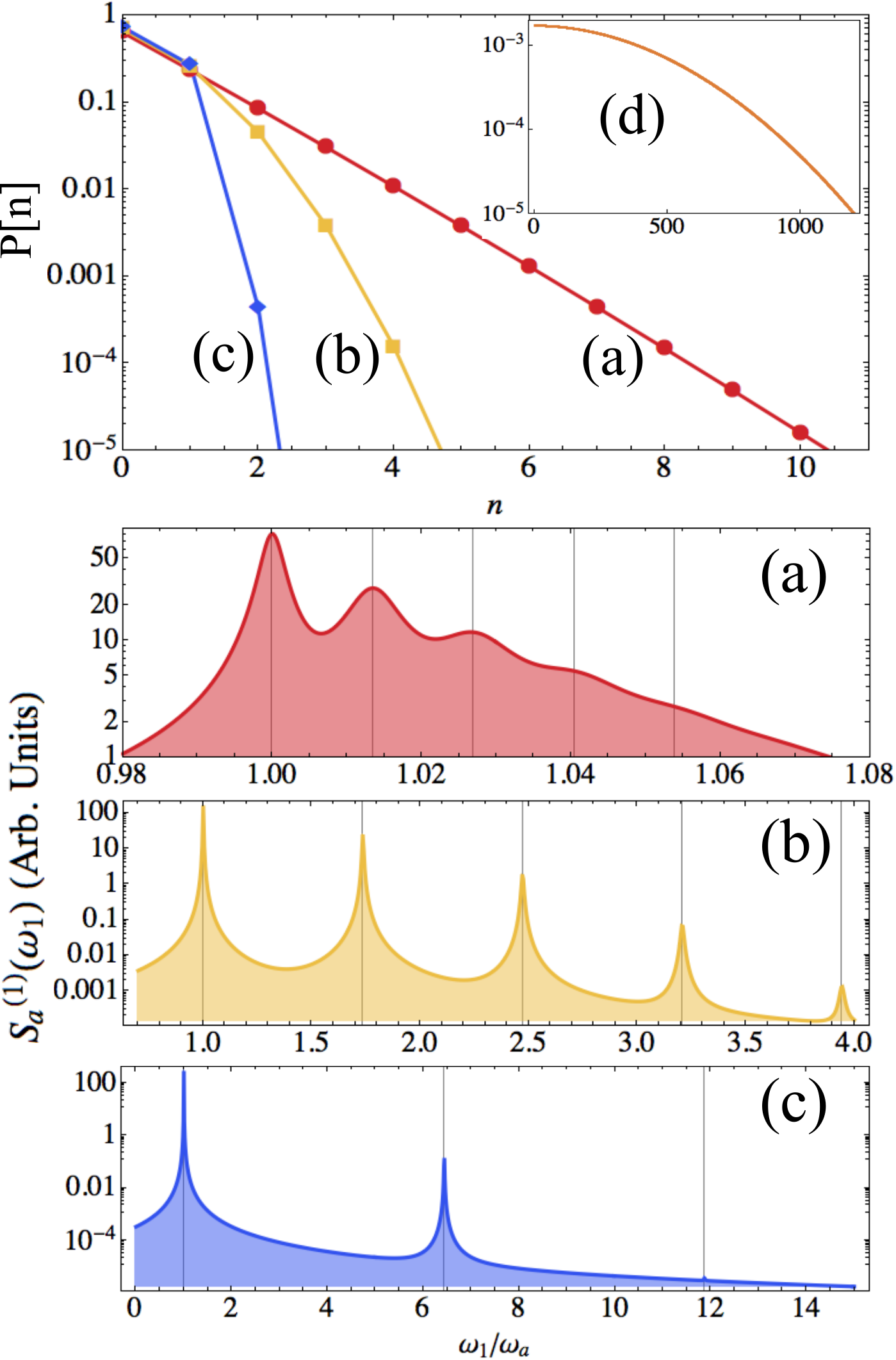}
  \caption{(color online) Photon number distribution for the four
    cases marked in Fig.~\ref{fig:kerrg2} with letters and the
    corresponding spectra of emission. Parameters are
    $\gamma_a=0.001\omega_a$, $T=\omega_a$, $\Gamma_1=0$ with (a)
    $U=e^{-5}\omega_a$, (b) $U=e^{-1}\omega_a$, (c)
    $U=e^{2}\omega_a$. (d) $T=e^{10}\omega_a$, $U=e^{-3}\omega_a$. The
    emission in case (d), not shown, is a large broad peak due to the
    large temperature-induced decoherence.}
  \label{fig:stats}
\end{figure}

The thermalized state achieved with the standard Kerr-nonlinearity
Hamiltonian $H_K$ should on physical grounds be given by the canonical
ensemble,
\begin{equation}
  \label{eq:WedFeb6174633CET2013}
  \rho_T=\frac{1}{Z}\sum_{n} e^{-\frac{\omega_a n+Un(n-1)}{T}}\ket{n}\bra{n}
\end{equation}
with $Z=\sum_{n} e^{-\frac{\omega_a n+Un(n-1)}{ T}}$ \cite{Alicki}. A
fundamental difference between Eqs.~(\ref{eq:FriApr19142616CEST2013})
and (\ref{eq:WedFeb6174633CET2013}) is that the first one has particle
statistics that are independent of the parameters of the Hamiltonian
and even independent of temperature with $g_a^{(N)}=\mean{(a^\dagger
  )^N a^N}/\mean{a^\dagger a}^N=N!$ while the latter one has particle
statistics depending on $T$ and $U$, including subpoissonian regions
with $g_a^{(2)}<1$ \cite{Quattropani}. We have plotted $g_a^{(2)}$
according to Eq.~(\ref{eq:WedFeb6174633CET2013}) in
Fig.~\ref{fig:kerrg2} as an illustration of the rich statistics that
the nonlinearity $U$ brings. Only in the region $U\ll \omega_a$ (case
(a)) do we recover statistics of thermal light fields with
$g_a^{(2)}=2$ while in the opposite regime (case (c)) we recover the
two-level system limit $g_a^{(2)}=0$ as expected when levels with more
than one particle $n>1$ are so high in energy that they cannot be
occupied by thermal fluctuations. The corresponding
photon-distribution functions, P$[n]=\bra{n}\rho_T\ket{n}$, are
plotted in the upper part of Fig.~\ref{fig:stats}, cases (a) and (c)
respectively.

It is interesting that the Kerr nonlinearity $H_K$ allows for
analytical solutions at the low and high temperature limits, as shown
in Fig.~\ref{fig:kerrg2} with separate upper and lower panels. In the
high occupation regime, $T\gg \omega_a$, mean values of any observable
can be obtained by transforming the sums over the number of
excitations, $n$, into an integral (continuous variable approximation)
giving for instance,
\begin{equation}
  \label{eq:WedFeb6174639CET2013}
  \mean{a^\dagger a}=\frac{1}{2}-\frac{\omega_a}{2U}+\frac{\sqrt{\frac{T}{\pi U}} e^{-\frac{(\omega_a-U)^2}{4TU}}}{1+\mathrm{Erf}(\frac{U-\omega_a}{2\sqrt{TU}})} \,, 
\end{equation}
with a limiting value of $\lim_{T\rightarrow
  \infty}g^{(2)}=\pi/2$. The corresponding photon-distribution
function, case (d), is plotted in the inset of the upper panel in
Fig.~\ref{fig:stats}. In the low occupation regime, at temperatures
$T<0.3\omega_a$, mean values can be obtained by truncating the sums in
the excitation number at $n=2$. From this, we can determine
analytically the non-linearity for which the statistics become
subpoissonian, $g^{(2)}\leq 1$, as,
\begin{equation}
  \label{eq:WedFeb6182356CET2013}
  U\geq \frac{T}{2}\mathrm{Ln} \big( e^{\omega_a/T}-1+\sqrt{e^{2\omega_a/T}-2e^{\omega_a/T}-1} \big) -\frac{\omega_a}{2}\,.
\end{equation}

In order to look into dynamical observables such as the transient
dynamics, $\rho(t)$, towards the thermalized steady state or the
spectrum of emission, which depend on $\gamma_a$ as well, we need the
correct master equation for the Hamiltonian $H_K$. Owing to the regime
of deep anharmonicity, the standard quantum optical master equation
with $\mathcal{\tilde L}_{T}$ would give a false description of the
dynamics. Indeed, this Lindblad dissipator is obtained in an optical
regime where the energy differences between subsequent levels are
almost the same, and in this case the mean excitation number in the
bath, i.e. the feeding factor, is almost the same for each transition
and thus fixed as a constant. In the spirit of Ref.~\cite{Blais}, one
can perform a perturbative expansion in the system-bath coupling
strength in the basis of the eigenstates $|j\rangle$ of the exact
Hamiltonian at hand, $H_{K}$ in this case, in order to derive the
Redfield equations~\cite{breuer} that describe the dissipative
processes. In our notation, we label the states $|j\rangle$ such that
$\omega_k > \omega_j$ as $k > j$. After some algebra, we obtain a
master equation with
\begin{eqnarray}\label{liouvillian} \nonumber
  \mathcal{L}_T&=& \sum_{j,k>j}\Gamma^{j k}_{a}
  \Big[ 1+\bar{n}_T(\Delta_{k j}) \Big]\mathcal{D}_{|j \rangle \langle
    k|} \\
  &+& \sum_{j,k>j}\Gamma^{j k}_{a}
  \bar{n}_T(\Delta_{k j}) \mathcal{D}_{|k \rangle \langle j|}\,.
\end{eqnarray}
In particular, in Eq. (\ref{liouvillian}), $\mathcal{D}$ operates on
the transition operators $|k \rangle \langle j|$ between the
\emph{k-}th and \emph{j-}th eigenstates. The relaxation coefficients
$\Gamma^{j k}_{a} = 2\pi d_{a}(\Delta_{k j}) \alpha^{2}_{a}(\Delta_{k
  j})| C^{a}_{j k}|^2$ can be interpreted as the full width at half
maximum of each $|k \rangle\rightarrow | j \rangle$ transition at zero
temperature, and they depend on the spectral density of the bath
$d_{a}(\Delta_{k j})$ and the strength of the coupling to the bath
$\alpha_{a}(\Delta_{k j})$ at their respective transition frequency
$\Delta_{k j} = \omega_{k} - \omega_{j}$, as well as on the transition
coefficients $C_{j k} = \langle j |(a + a^{\dagger})| k \rangle$. For
a flat spectral density $d_{a}(\Delta_{k j})$, and couplings
$\alpha_{a}(\Delta_{k j})$ that are frequency independent (Markov
approximation), the relaxation coefficients reduce to $ \Gamma^{j
  k}_{a} = \gamma_{a}|C^{a}_{j k}|^2$, where $\gamma_{a}$ is the
standard damping rate. For the Hamiltonian $H_K$, the eigenstates
remain the number states $\ket{n}$ and the energy difference between
them is $\Delta \epsilon_n $. In order to solve Eq. (\ref{master-eq})
in the steady state, we first put the density matrix elements
$\bra{j}\rho\ket{k}$ in a vector that we denote $\mathbf{v}$ and
rewrite the master equation in matricial form,
\begin{equation}
  \label{eq:TueMar20012758CET2012}
  \partial_\tau \mathbf{v}(\tau) = M \mathbf{v}(\tau)\,.
\end{equation}
The transient solution $\mathbf{v}(\tau)=e^{M\tau}\mathbf{v}(0)$
converges to the steady state in the long time limit as
\begin{equation}
  \label{eq:TueMar20013537CET2012}  
  \mathbf{v}^\mathrm{ss}=\lim_{\tau\rightarrow
    \infty}\mathbf{v}(\tau)=\lim_{\tau\rightarrow
    \infty}e^{M\tau}\left( \begin{array}{c}
      1 \\
      0 \\
      \vdots
    \end{array}
  \right)\,,
\end{equation}
where we have chosen the vacuum as the initial condition. Since we
employ the justified assumption of a unique steady state
\cite{Spohn,Schirmer}, the initial state is irrelevant and all the
relevant information is encoded in $e^{M\tau}$.  With this, we arrive
to the thermal steady state defined by
Eq.~(\ref{eq:WedFeb6174633CET2013}).

From the dynamics, we additionally obtain the spectrum of emission in
the steady state (set at $t=0$),
\begin{equation}
  \label{eq:spe-QRF}
  S_a^{(1)}(\omega_1)=\frac{1}{\pi}\Re\int_0^{\infty} d\tau e^{-\frac{\Gamma_1}{2}\tau}e^{-i\omega_1\tau} \mean{ a^\dagger (0)a(\tau)}\,,
\end{equation}
using the quantum regression formula~\cite{delvallebook10} to obtain
the two-time correlator $\mean{ a^\dagger (0)a(\tau)}$ from the master
equation~(\ref{eq:TueMar20012758CET2012}). The linewidth $\Gamma_1$
provides the uncertainty in the frequency detection of the measurement
apparatus~\cite{eberly77a}. According to Ref.~\cite{delvalle12a}, the
spectrum can also be computed as the steady state population of an
output detector or \emph{sensor} with central frequency $\omega_1$
which is very weakly coupled to the measured field $a$,
\begin{equation}
  \label{eq:S1}
  S_a^{(1)}(\omega_1)\propto \mean{n_1}\,,
\end{equation}
where $\mean{n_1}$ is the sensor occupation and $\Gamma_1$ its decay
rate. In the Appendix we explain in detail the method to compute the
spectrum within the density matrix formalism, providing a
semi-analytical formula in terms of the matrix $M$,
Eq.~(\ref{eq:WedMar21032339CET2012final}).

Fig.~\ref{fig:stats} shows three examples of spectra of emission
together with the photon number distribution for a fixed temperature
and three different nonlinearities, marked in Fig.~\ref{fig:kerrg2} as
(a), (b) and (c). The photon number distribution would be that of (a)
for the three cases, had we solved the master equation with
Liouvillian $\mathcal{\tilde L}_T$ given in
Eq.~(\ref{eq:wrong-thermal}). The proper Liouvillian,
Eq.~(\ref{liouvillian}), gives rise to a distribution that is
$U$-dependent as required on physical grounds. The spectrum of
emission is different in all three cases even under $\mathcal{\tilde
  L}_T$, but the intensities of the peaks are not accurately
obtained. As expected, increasing the nonlinearity separates the
different peaks, produced in the different transitions between
subsequent energy levels, and makes it increasingly harder to populate
high energy levels. At very large $U$, only the peak at $\omega_a$
survives as it corresponds to the emission of a two-level system.

\section{Full anharmonic Hamiltonian}
\label{sec:full}

\begin{figure}
  \includegraphics[width=\linewidth]{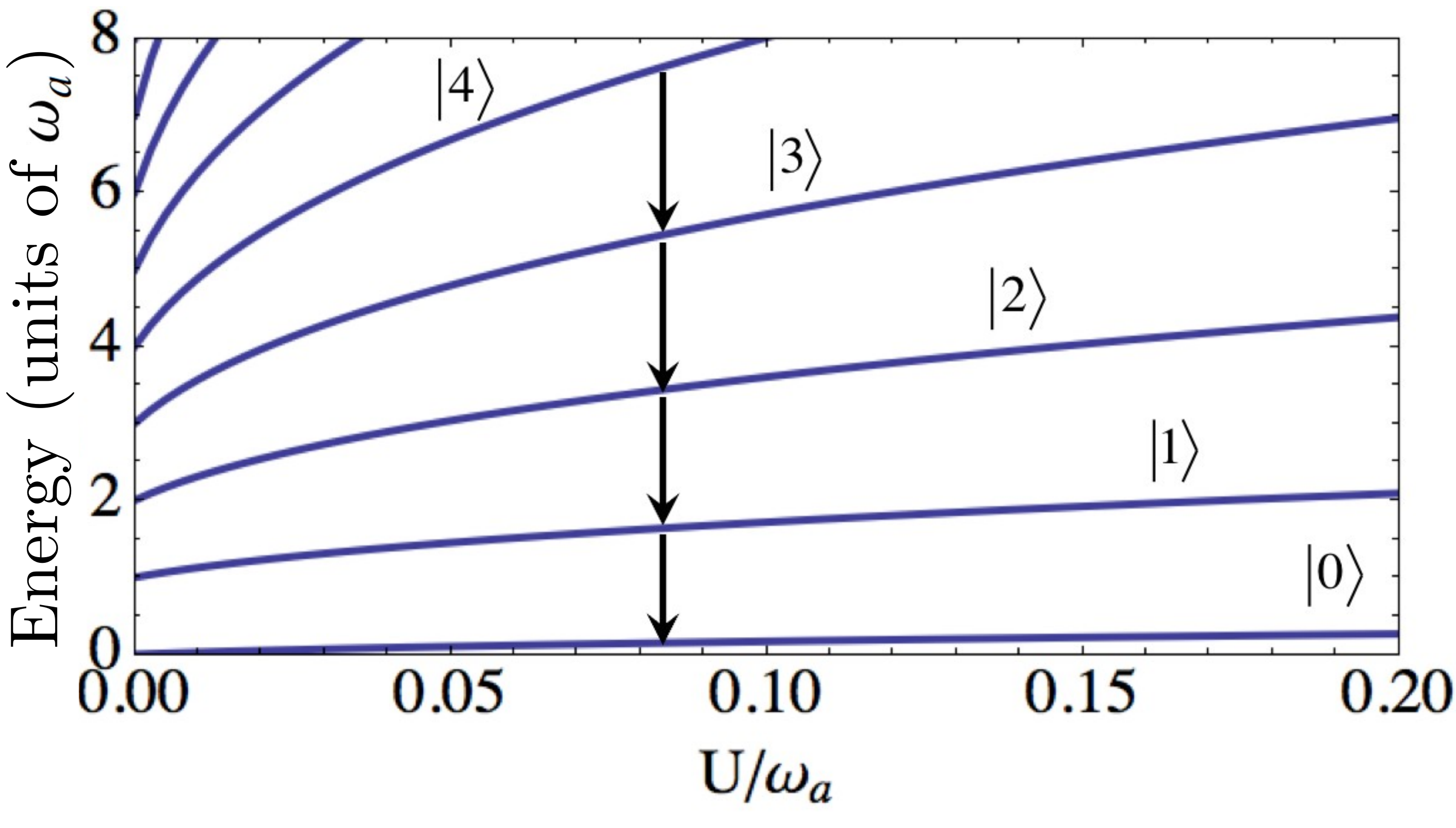}
  \caption{Energy levels of $H_{S}$ as a function of the nonlinearity
    $U$ for repulsive nonlinearities $U>0$.}
  \label{fig:levels} 
\end{figure}

\begin{figure}
 \includegraphics[width=\linewidth]{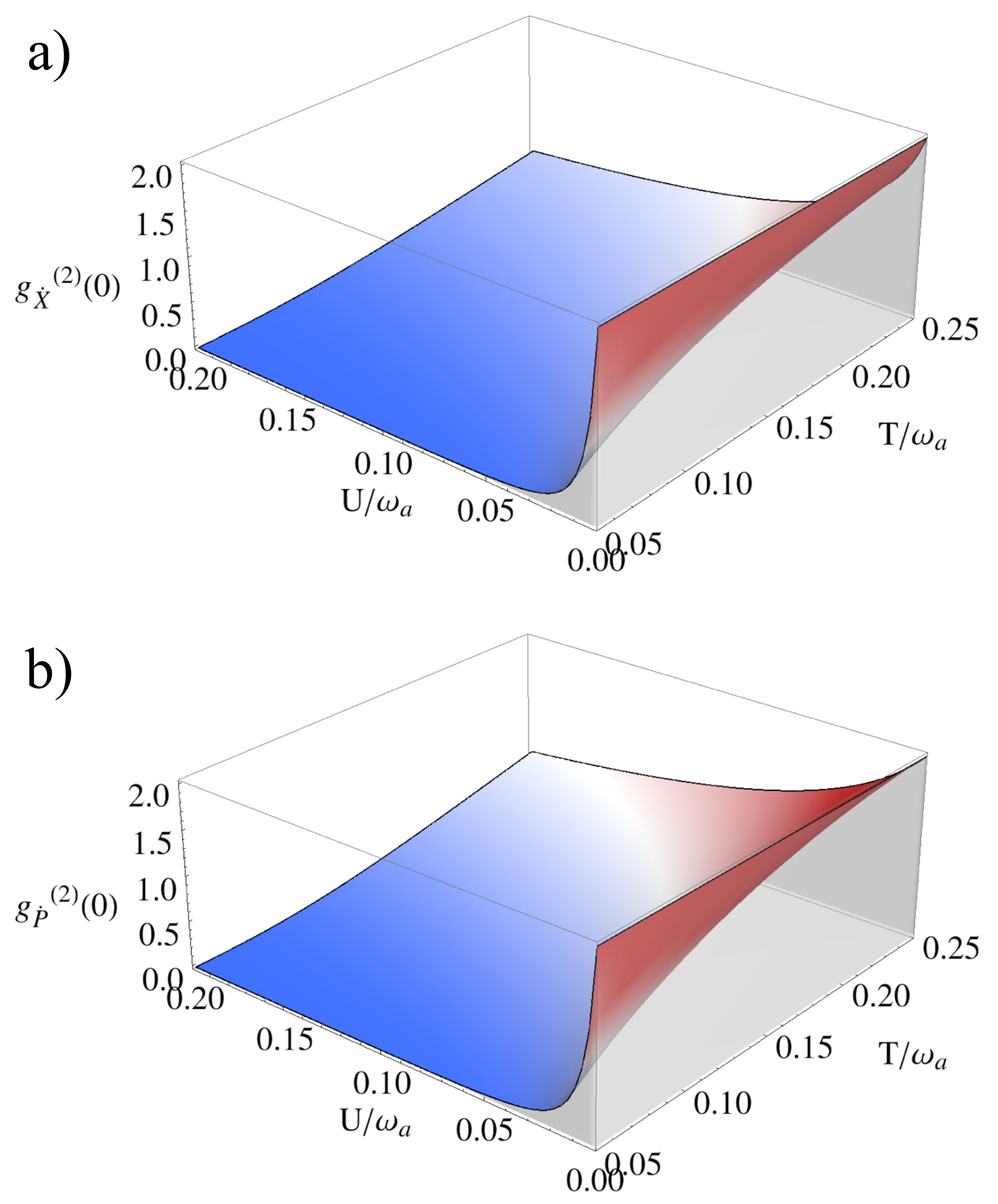}
 \caption{(color online) Zero-delay second order correlation function,
   $g^{(2)}(0)$, at thermal equilibrium, calculated separating
   positive and negative frequency components with respect to $H_{S}$
   for the $\dot{X}$ quadrature (plot a), i.e.  $\langle
   \dot{X}^{-}\dot{X}^{-}\dot{X}^{+}\dot{X}^{+} \rangle / \langle
   \dot{X}^{-} \dot{X}^{+} \rangle^2$, and for the $\dot{P}$ quadrature (pot b).}
\label{fig:spectrum1}
\end{figure}

For higher photon numbers, a more accurate description of the
\emph{ultra-anharmonic} regime is provided by the
Hamiltonian $H_{S}$. After the diagonalization of $H_{S}$,
whose eigenstates are no longer the number states, the steady state
density matrix of the canonical ensemble reduces to,
\begin{equation} 
  \label{eq:canonical} 
  \rho_{T} =\frac{1}{Z} \sum_j e^{-\epsilon_j/T} \ket{j}\bra{j}
\end{equation}
where $\epsilon_{j}$ is the \emph{j-}th eigenvalue of $H_{S}$ and $Z =
\sum_j e^{-\epsilon_j/T}$. These eigenenergies are plotted in
Fig.~\ref{fig:levels} as a function of the nonlinearity.  In this
case, not only the dissipation term in the master equation but also
the photodetection has to be modified in order to correctly describe
the nonlinearity of the system~\cite{RidolfoPrl2012}. Otherwise,
unphysical results are found such as a stream of output photons when
the system is in its ground state \cite{Werlang}.  By following the
original photodetection formulation by Glauber, the probability per
second that a photon is absorbed by an ideal detector is proportional
to $\langle E^-(t) E^+(t) \rangle$, where $E^\pm(t)$ are the positive
and negative frequency components of the electric field operator of
the output. In the same way, the photon correlation functions are
straightforwardly calculated as \cite{Milonni,SavastaPRA1996} $\langle
E^-(t) E^-(t') E^+(t') E^+(t) \rangle$, with all the positive
frequency operators to the right and all the negative frequency
operators to the left. Following Ref.~\cite{RidolfoPrl2012}, by
expressing the cavity electric-field operator in the eigenbasis, we
derive correlation functions for the output fields which are valid for
an arbitrary nonlinearity. Let us define the quadrature operators $X =
X_{0}(a + a^{\dagger})$, and its conjugate momentum $P = -i P_{0}(a -
a^{\dagger})$, with their time derivatives, $\dot{X} = i [H , X]$ and
$\dot{P} = i [H , P]$.  The input-output relations can be derived in a
very general way~\cite{GardinerZoller}, and for an $X$ quadrature
coupled to the electric field of the output channel, one finds,
\begin{equation}\label{in-out}
  E_{\text{out}} = E_{\text{in}} - \sqrt{\kappa} \, \dot{X}
\end{equation}
where $\kappa$ is the associated decay rate into the output channel.
Likewise, Eq. (\ref{in-out}) can be generalized for the $P$ quadrature
just replacing $X$ with $P$. Although this latter replacement seems to
be harmless, it is worthwhile to notice that it has crucial
significance in terms of physical observables. In fact, the output
field has to reflect the symmetries of the system as it is explained
further on. For input fields in the vacuum state, we define the
delayed second order correlation function as
\begin{equation} 
  g_{\dot{X}}^{(2)}(\tau) = \lim_{t \to \infty} \frac{\langle
    \dot{X}^{-}(t) \dot{X}^{-}(t+\tau) \dot{X}^{+}(t+\tau)
    \dot{X}^{+}(t) \rangle}{\langle \dot{X}^{-}(t) \dot{X}^{+}(t)
    \rangle^{2}}. 
\end{equation}
Thus, obtaining the photon correlations for the output fields requires
calculating the positive and negative frequency components of the
operator $\dot{X}$, namely $\dot{X}^{+}$ and
$\dot{X}^{-}$~\cite{RidolfoPrl2012}. By expanding $\dot{X}$ in the
basis of energy eigenstates $|j\rangle$, it is easy to find $\dot{X}^+
= -i \sum_{j,k>j} \Delta_{kj}X_{jk} | j \rangle \langle k |$, where
$X_{jk} = \langle j | X | k \rangle$ and $\dot{X}^{-} =
(\dot{X}^{+})^\dag$. In Fig.~\ref{fig:spectrum1} we plot the second
order coherence function at zero delay for both quadratures, $X$ and
$P$. We observe similar behaviors with thermal and antibunched regions
for small and large nonlinearities respectively. These functions are
always bounded between 0 and 2 as in Fig.~\ref{fig:kerrg2} and thus
show physically meaningful results.
%

\begin{figure}
  \includegraphics[width=\linewidth]{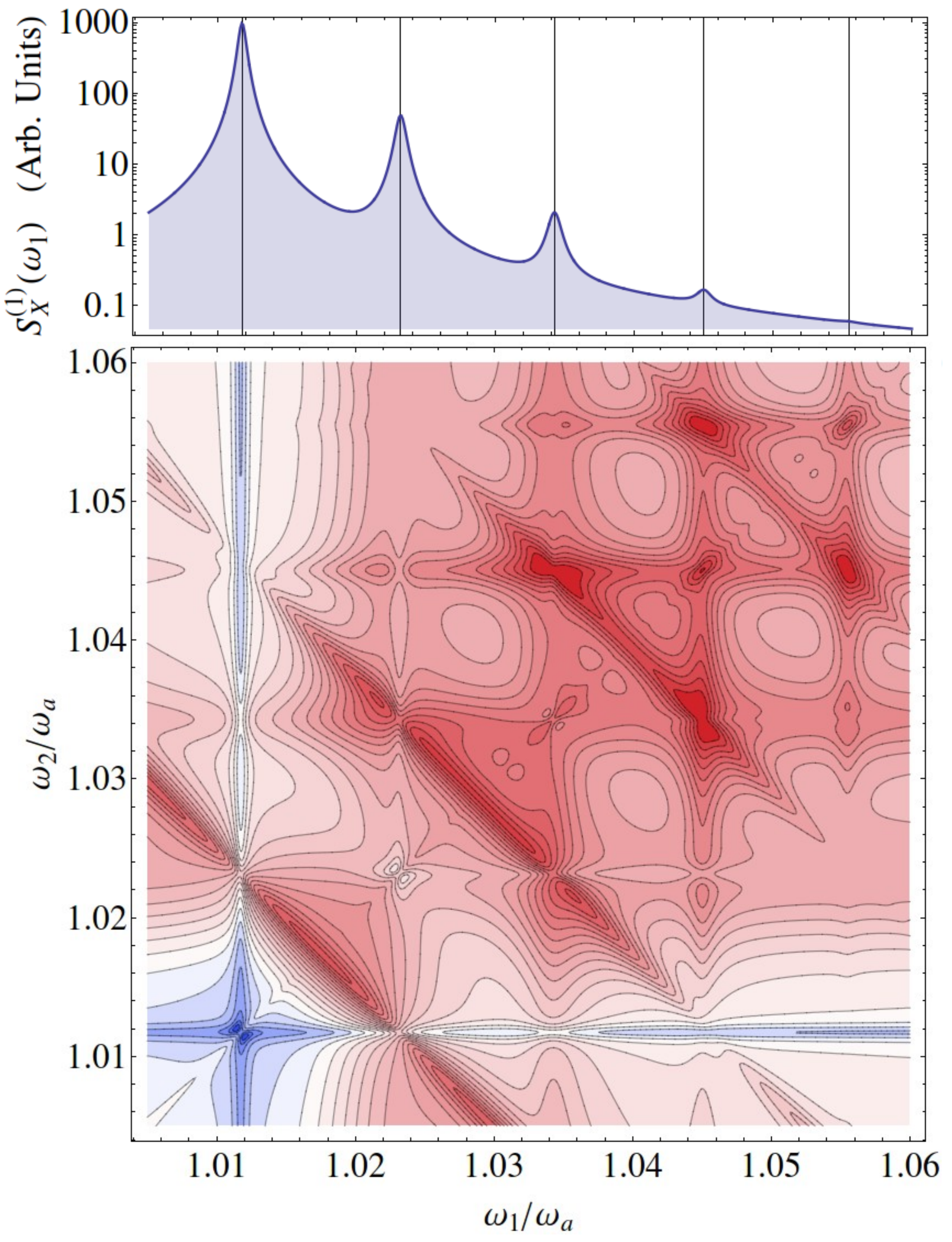}
  \caption{(color online) One-photon spectrum (top panel), and two
    photon spectrum at zero-delay time (bottom panel) at thermal
    equilibrium, calculated from
    Eqs.~(\ref{eq:WedMar21032339CET2012final}) and
    (\ref{eq:WedMar21195020CET2012}). Parameters are: $U = 10^{-3}
    \omega_{a}$, $T = 0.3 \omega_{a}$, $\gamma_{a} =
    10^{-4}\omega_{a}$, $\Gamma_1=\Gamma_2= 5\times10^{-4}
    \omega_{a}$. The vertical grid lines mark the positions of the
    transitions in the system. The color scale ranks from the minimum
    to maximum value: 0.063 (darkest) blue, 1 white, 1572 (darkest)
    red.}
  \label{fig:spectrum} 
\end{figure}

Let us finally turn into steady state spectral functions for one and
two photons.  Here, we couple the sensors to the $X$ operator, and we
easily find the input-output relations for the electric field of the
sensors as in Eq.~(\ref{in-out}). Since the sensors are very weakly
coupled to the system, we can describe them as a single mode
resonance, and the derivative of these operators reduces to
$\dot{\varsigma_i} = -i \omega_{i} \varsigma_i$, being $\omega_{i}$
the frequency of the \emph{i-}th sensor.  Then, the output electric
field is just,
\begin{equation}\label{in-out2}
  E_{\text{out}} = E_{\text{in}} -i \, \epsilon_{i} \, \omega_{i}  \, X
\end{equation}
where $\epsilon_{i}$ is the coupling strength between the \emph{i-}th sensor and the oscillator.
Thus the power spectrum reads,
\begin{equation}
  \label{eq:SatJul14173450CEST2012}
  S_{X}^{(1)}(\omega_1)=\frac{\omega_1^2}{\pi}\Re\int_0^{\infty}
  d\tau
  e^{-\frac{\Gamma_1}{2}\tau}e^{-i\omega_1\tau}\langle
  X^-(0)X^+(\tau)\rangle \,,
\end{equation}
and the normalized two-photon spectrum of emission, computed as the
cross intensity-intensity correlations between two sensors with
frequencies $\omega_1$ and $\omega_2$ read,
\begin{equation}
  \label{eq:FriJan18161900CET20132}
  g_{X}^{(2)}(\omega_1;\omega_2)=\frac{  S_{X}^{(2)}(\omega_1;\omega_2)}{S_{X}^{(1)}(\omega_1) S_{X}^{(1)}(\omega_2)}=\frac{\mean{n_1n_2}}{\mean{n_1}\mean{n_2}}
  \,.
\end{equation}
In the Appendix we derive semi-analytical expressions for both the one
and two-photon spectrum in the steady state as a function of the
master equation coefficient matrix of the system only, $M$,
Eq.~(\ref{eq:WedMar21195020CET2012}). We plot both
$S_{X}^{(1)}(\omega_1)$ and $g_{X}^{(2)}(\omega_1;\omega_2)$ in
Fig.~\ref{fig:spectrum} for the full Hamiltonian $H_S$ in a region
where the mean number of excitations is $\mean{X^-X^+} = 0.035$ and
the total second order coherence function is very close to thermal,
$g_{\dot{X}}^{(2)}(0) = 1.943$. The one-photon spectrum provides again
the transition energies in the system and their frequency uncertainty
(once deconvoluted from the detector precision $\Gamma_1$). The
two-photon spectrum provides a clear picture of the level
structure~\cite{gonzaleztudela13a}. First, we observe the
characteristic blue butterfly shape around each transition frequency,
$\omega_1=\omega_2=\Delta_{j+1 j}$, as they are isolated from the rest
by the nonlinearity. This is specially visible for the single
excitation to ground state transition,
$\omega_1=\omega_2=\Delta_{10}\approx \omega_a$, where antibunching is
strong (in deep blue color). At larger nonlinearities $U$, this is the
only remaining feature as it corresponds to the two-level
system. Second, we observe the cascade type of correlations for every
pair of consecutive transition frequencies, $\omega_1=\Delta_{j+1 1}$,
$\omega_2=\Delta_{j j-1}$. This is recognized by a dip in the
correlations, as compared to the antidiagonal lines that cross these
points, getting close to one. This is the middle value for $g^{(2)}$
at $\tau=0$, between the bunching effect when the sign of the delay
follows the natural cascade order (first $\Delta_{j+1 1}$ and then
$\Delta_{j j-1}$) and the opposite delay sign that produces an
antibunching effect. Finally, the diagonal and antidiagonal patterns
are filtering induced effects. The diagonal line corresponds to an
extra bunching produced by measuring indistinguishable photons,
$\omega_1=\omega_2$, as explained in detail in
Refs.~\cite{delvalle12a} and \cite{gonzaleztudela13a}. The
antidiagonal lines, given by
$\omega_1+\omega_2=\epsilon_j-\epsilon_{j-2}$ for $j\geq 2$,
correspond to \emph{leapfrog} processes, where two-photons are emitted
at the same time (within the time uncertainty window $1/\Gamma_1$)
without populating the intermediate level. The nonlinearity allows
these antidiagonal lines to split and be individually resolved,
opening, therefore, the possibility of two-photon state generation in
the system~\cite{delvalle13a}.

\section{Attractive Nonlinearities}
\label{sec:attractive}

Circuit quantum electrodynamics is a prime candidate for the
realization of attractive ($U<0$) ultra strong optical nonlinearities
of the form we investigate here.  We thus complete our discussion with
results for an attractive nonlinearity, $U < 0$, as it is for example
realized in transmon qubits~\cite{Koch07} or in a transmission line
resonator where the central conductor is intersected by a dc
SQUID~\cite{Leib12,Bourassa12}.

A version of the Hamiltonian (\ref{h-full}) can be implemented for
microwave fields in circuit quantum electrodynamics where the
nonlinearity is provided by a Josephson junction.  The associated
nonlinear inductance is described by a term $E_{J} \cos \phi$ in the
Hamiltonian, which yields the Hamiltonian (\ref{h-full}) by
identifying $\phi = \sqrt{2 E_{C}/E_{J}} \left(a +
  a^{\dagger}\right)$, $\omega_{a} = \sqrt{8 E_{C} E_{J}}$ and $U_{2
  n} = - E_{J} \left(-2 E_{C}/E_{J}\right)^{n}/(2n)!$~\cite{Leib12},
where $E_{C}$ and $E_{J}$ are the charging and Josephson energies of
the considered circuit.  Here, $U = - E_{C}/12$ is negative and the
power series in Eq.~(\ref{h-full}) can only be truncated for
sufficiently small ratios $E_{C}/E_{J}$ as for example in a
transmon~\cite{Koch07}.

\begin{figure}
  \includegraphics[width=\linewidth]{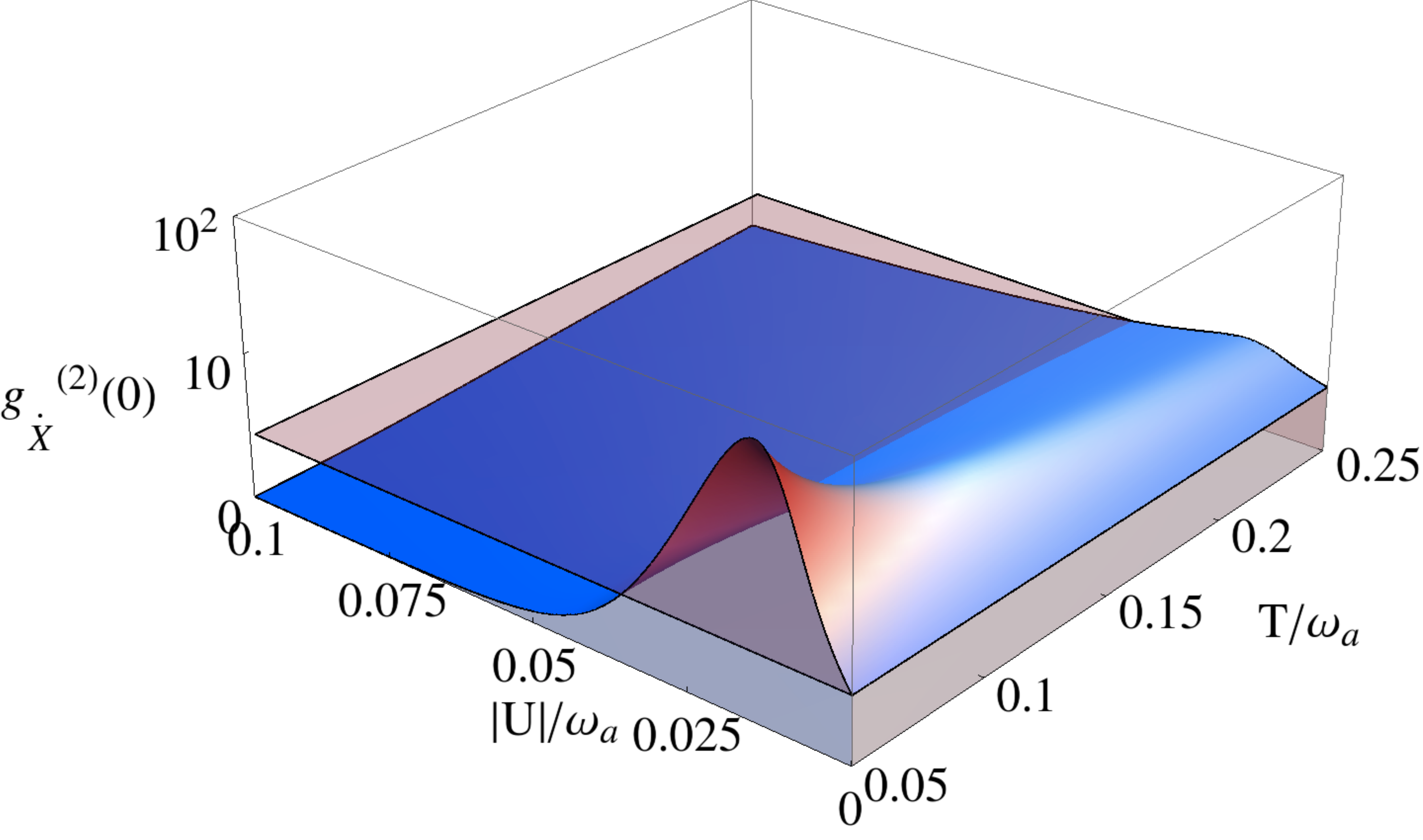}
  \caption{(color online) $g_{\dot{X}}^{(2)}(0)$ as a function of
    temperature and nonlinearity for the Hamiltonian $H_S$ (including
    the corresponding $U_{6}$-term) with $U<0$,
    c.f. Eq.~(\ref{hamiltonian1}).  For comparison we show the plane
    $g_{\dot{X}}^{(2)} (0)= 2$ for standard thermal particle
    statistics of a non-interacting field.}
  \label{fig:josephson} 
\end{figure}

Fig.~\ref{fig:josephson} shows $g_{\dot{X}}^{(2)}(0)$ according to
Eq.~(\ref{eq:WedFeb6174633CET2013}) for the case where $U < 0$.  For
comparison we also show the plane $g_{\dot{X}}^{(2)}(0) = 2$ for
standard thermal particle statistics of a non-interacting field. One
can clearly identify a region with enhanced bunching
$g_{\dot{X}}^{(2)}(0) > 2$ for moderate but nonzero interactions and
low temperatures. For stronger interactions $|U|/\omega_{a} \gtrsim
0.04$ the field becomes strongly antibunched. Note that we have kept
here the next order $U_{6}$ of the nonlinearity to ensure that the
spectrum of $H_{S}$ always has a lower bound.

\begin{figure}
  \includegraphics[width=\linewidth]{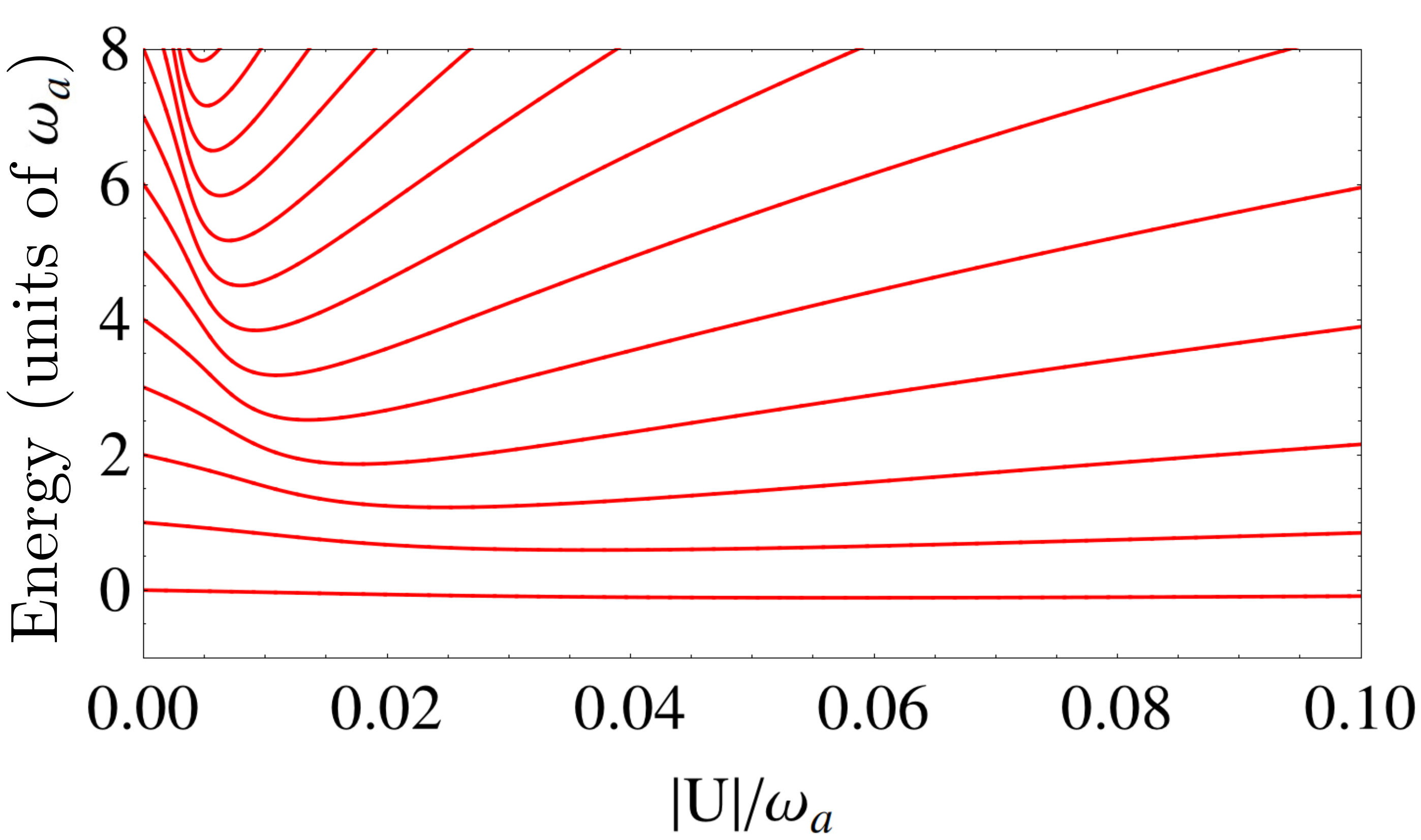}
  \caption{Energy levels of $H_{S}$ (including the corresponding
    $U_{6}$-term) as a function of the nonlinearity $U$ for attractive
    nonlinearities $U<0$.}
  \label{fig:josephsonlevels} 
\end{figure}

These features can be well understood by inspection of the energy
levels of $H_{S}$ as a function of the nonlinearity $|U|$, see
Fig.~\ref{fig:josephsonlevels}.  In the parameter region where
bunching appears, the transition energy between the first and second
excited state is smaller than between ground and first excited state.

\section{Conclusions}
\label{sec:concl}

We have analyzed the effect of a nonlinearity, $U$, that can be as
large as the natural frequency, $\omega_a$, on the thermal equilibrium
properties of a single mode. We have considered two types of quartic
nonlinearities in the Hamiltonian of the system and derived the
adequate master equation for the time evolution in contact with the
thermal bath, as well as the output fields that can be measured in
each case. In order to obtain a physical solution in agreement with
the canonical ensemble, the Lindblad forms that describe dissipation
and excitation must be in terms of the new eigenstates of the
Hamiltonian, which are obtained numerically beforehand. We have
focused on spectral and statistical properties of the system in a
separated (one-photon or power spectrum of emission and second order
coherence function) and combined (frequency resolved second order
correlations or two-photon spectrum) way. We have derived a
semi-analytical expression for the last one, following a sensor
approach, only in terms of the master equation coefficients and the
steady state density matrix. These observables offer complementary
information about the different regimes appearing in the system when
varying $U/\omega_a$ and $T/\omega_a$. At small nonlinearities the
mode is in a thermal state, $g^{(2)}=2$, and one can apply the
standard approximations for the Lindblad terms and output field. At
large nonlinearities, however, the behavior is effectively close to
that of a two-level system, with antibunched statistics, $g^{(2)}=0$,
and a single transition isolated in energy. In the intermediate
regimes a cascade of well defined transitions occurs providing a set
of peaks in the spectrum and $0<g^{(2)}<2$. For attractive
nonlinearities, $U<0$, low temperatures and small interactions,
bunching can be enhanced above the thermal value, $g^{(2)}>2$.

\begin{acknowledgments}
  AR acknowledges support from the Emmy Noether project HA 5593/1-1
  (DFG), EdV from the Alexander von Humboldt Foundation and CAM under
  contract S2009/ESP-1503, and MJH from the Emmy Noether project HA
  5593/1-1 and CRC 631 (DFG).
\end{acknowledgments}

\appendix

\section{Derivation and formulas for one- and two-photon spectrum}

Following the general formalism in Ref.~\cite{delvalle12a}, we derive
semi-analytical expressions for the steady state one- and two-photon
spectra of emission from a master equation approach, in an analogous
way as done in Ref.~\cite{gonzaleztudela13a}, in terms of the relevant
correlators in the system and their equations. We call the measured
annihilation and creation field operators $X^+$ and $X^-$.

We define two reordering matrices, $T_\pm$, which, when acting on
$\mathbf{v}$, substitute each element in it, $\bra{m}\rho\ket{n}$, by
$\bra{m}X^+\rho\ket{n}$ for $T_+$ and $\bra{m}\rho X^-\ket{n}$ for
$T_-$.  These matrices always exist, in infinite or in truncated
Hilbert spaces (where, if truncation is to order~$n_\mathrm{max}$, we
set $\bra{n_\mathrm{max}}X^+\rho\ket{n}=0$ and $\bra{n}\rho
X^-\ket{n_\mathrm{max}}=0$ for all $n$).

We now consider two sensors (although this can be generalized to an
arbitrary number) with operators $\varsigma_i$, $i=1,2$ and
linewidths~$\Gamma_i$ coupled to the system with
strength~$\varepsilon_i$ such that the dynamics of the system is
probed but is otherwise left unperturbed. This requires the tunnelling
rates $\varepsilon_i$ to fulfill
$\varepsilon_i\ll\sqrt{\Gamma_i\gamma_Q/2}$, where $\gamma_Q$ is the
smallest system decay rate.  The new density matrix that includes the
sensors, $\rho_\mathrm{sen}$, follows a modified master equation where
the photonic tunnelling terms,
$H_\mathrm{sen}=\sum_{i=1}^N[\omega_i\ud{\varsigma_i}\varsigma_i +
\varepsilon_i (X^+ \ud{\varsigma_i}+X^-\varsigma_i)]$, are added to
the original Hamiltonian, and the sensor decay terms
$\sum_{i=1}^N\Gamma_i\mathcal{D}_{\varsigma_i} \rho_\mathrm{sen}$ are
added to the dissipative part.

We define new vectors, $\mathbf{w}$, each of them containing the
system density matrix (in the same order as $\mathbf{v}$) but for a
given combination of sensor states. That is,
$\mathbf{w}[\mu_1\nu_1][\mu_2\nu_2]$ contains elements
$\bra{j,\mu_1,\mu_2}\rho_\mathrm{sen}\ket{k,\nu_1,\nu_2}$ with
$j,k=1,2,\ldots$ labelling the system eigenstates. The sensors are
two-level systems so the indices $\mu_i$ and $\nu_i$ take the values 0
or 1. The reduced system density matrix is recovered tracing over the
sensors as
$\mathbf{v}=\sum_{\mu_1=0,1}\sum_{\mu_2=0,1}\mathbf{w}[\mu_1\mu_1][\mu_2\mu_2]$.
The reduced sensor density matrix, is obtained by tracing over the
system as $u[\mu_1\nu_1][\mu_2\nu_2]=\sum_m
\bra{m,\mu_1,\mu_2}\rho_\mathrm{sen}\ket{m,\nu_1,\nu_2}=\Tr_\mathrm{sys}(\mathbf{w}[\mu_1\mu_1][\mu_2\mu_2])$
(noting that $\mathbf{w}$ is a vector so tracing means reconstructing
it in a matrix form first). Let us also note that
$\mathbf{w}[\mu_1\nu_1]=\sum_{\mu_2=0,1}\mathbf{w}[\mu_1\nu_1][\mu_2\mu_2]$
when we trace over the second sensor only.

The part of the master equation concerning each of the sensors and
their coupling to the system, reads,
\begin{widetext}
\begin{subequations}
  \label{eq:FriJan18130845CET2013}
  \begin{align}
    \partial_t
    \bra{m,\mu_1}\rho_\mathrm{sen}\ket{n,\nu_1}\Big|_\mathrm{sensor
      1}=& [(\nu_1-\mu_1)i\omega_1-(\mu_1+\nu_1){\frac{\Gamma_1}2}
    ]\bra{m,\mu_1}\rho_\mathrm{sen}\ket{n,\nu_1} \\
    &+\Gamma_1(1-\mu_1)(1-\nu_1) \bra{m,1}\rho_\mathrm{sen}\ket{n,1}\\
    +i\varepsilon_1 \Big[&
    -\mu_1\bra{m,0}X^+ \rho_\mathrm{sen}\ket{n,\nu_1}+\nu_1\bra{m,\mu_1}\rho_\mathrm{sen}X^-\ket{n,0}\\
    &-(1-\mu_1) \bra{m,1}X^-\rho_\mathrm{sen}\ket{n,\nu_1}+(1-\nu_1)
    \bra{m,\mu_1}\rho_\mathrm{sen}X^+\ket{n,1}\Big]\,.
  \end{align}
\end{subequations}
The sensors are mere spectators of the emission from the system and do
not alter its dynamics in any way. They are barely populated
($\langle\ud{\varsigma_i}\varsigma_i\rangle\ll 1$) and we can make the
approximation that their ground state provides the system steady state
(to second order in the couplings): $\mathbf{v}\approx
\mathbf{w}[00][00]$. In the same way, tracing over the state of one
sensor (for instance, the second) can be achieved by just fixing it in
its ground state: $\mathbf{w}[\mu_1\nu_1]\approx
\mathbf{w}[\mu_1\nu_1][00]$.

In order to obtain the equations of motion valid to leading order
in~$\varepsilon_{1,2}$, we note that the line b in
Eq.~(\ref{eq:FriJan18130845CET2013}) only applies to the element where
$\mu_1,\nu_1=0$, which is of no interest for us (and we know
corresponds to the steady state of the system anyway). We can drop
that line for our considerations. Furthermore, the last line d, can be
dropped as well because it links the element with $\mu_1 $ or
$\nu_1=0$ to $\mu_1$ or $\nu_1=1$.  This would lead to elements of
higher order in the couplings, which we discard. In physical terms,
line d corresponds to the process of the system absorbing an
excitation from the sensors (back action), which we
neglect. Therefore, we only keep lines a and c, obtaining, for the two
sensor vector,
\begin{multline}
  \label{eq:FriMar30165907CEST2012}
  \partial_t \mathbf{w}[\mu_1\nu_1][\mu_2\nu_2]= \large\{ M+[(\nu_1-\mu_1)i\omega_1-(\mu_1+\nu_1){\frac{\Gamma_1}2}
  +(\nu_2-\mu_2)i\omega_2-(\mu_2+\nu_2)\frac{\Gamma_2}2]\mathbf{1} \large\}
  \mathbf{w}[\mu_1\nu_1][\mu_2\nu_2]\\
  +\mu_1(-i\varepsilon_1T_+)\mathbf{w}[0\nu_1][\mu_2\nu_2]+\nu_1(i\varepsilon_1T_-)\mathbf{w}[\mu_10][\mu_2\nu_2]
  +\mu_2(-i\varepsilon_2T_+)\mathbf{w}[\mu_1\nu_1][0\nu_2]+\nu_2(i\varepsilon_2T_-)\mathbf{w}[\mu_1\nu_1][\mu_20]\,.
\end{multline}
The matrix $M$ contains the system dynamics. This is equivalent to
Eq. (12) of the supplemental material in Ref.~\cite{delvalle12a}. The
equations can be solved recursively,
  \begin{multline}
    \label{eq:FriMar30202508CEST2012}
    \mathbf{w}[\mu_1\nu_1][\mu_2\nu_2]=\frac{-1}{M+[(\nu_1-\mu_1)i\omega_1-(\mu_1+\nu_1)\frac{\Gamma_1}2+(\nu_2-\mu_2)i\omega_2-(\mu_2+\nu_2)\frac{\Gamma_2}2]\mathbf{1}}\\
    \times \Big\{
    \mu_1(-i\varepsilon_1T_+)\mathbf{w}[0\nu_1][\mu_2\nu_2]+\nu_1(i\varepsilon_1T_-)\mathbf{w}[\mu_10][\mu_2\nu_2]
    +\mu_2(-i\varepsilon_2T_+)\mathbf{w}[\mu_1\nu_1][0\nu_2]+\nu_2(i\varepsilon_2T_-)\mathbf{w}[\mu_1\nu_1][\mu_20]\Big\}
    \,.
  \end{multline}

\subsection{One-photon spectrum of emission (one sensor)}

The single-photon physical spectrum of the field $X$ is given in the
steady state (set at $t=0$) by Eq.~(\ref{eq:SatJul14173450CEST2012})
in the main text, that is, by the average population, in the steady
state, of any one of the two sensors, say, the first one,
\begin{equation}
  \label{eq:WedMar21033437CET20122}
  \mean{n_1}=\mean{\varsigma^\dagger_1\varsigma_1}=\Tr_\mathrm{sys}(\mathbf{w}[11][00])=\frac{\varepsilon_1^2}{\Gamma_1}(2\pi) S_{\Gamma_1}^{(1)}(\omega_1)\,,
\end{equation}
as was proven in Ref.~\cite{delvalle12a}. The approximated equation of
motion of such element, reads $\partial_t \mathbf{w}[11][00] =
(M-\Gamma_1\mathbf{1})\mathbf{w}[11][00]+(-i\varepsilon_1T_+)
\mathbf{w}[01][00] +(i\varepsilon_1T_-) \mathbf{w}[10][00]$, so we
have,
\begin{equation}
  \label{eq:WedMar21032339CET2012}
  \mathbf{w}[11][00]=\frac{-1}{M+(-\Gamma_1)\mathbf{1}}\Big\{ (-i\varepsilon_1T_+) \mathbf{w}[01][00]
  +(i\varepsilon_1T_-) \mathbf{w}[10][00] \Big\}\,.
\end{equation}
Using the solution Eq.~(\ref{eq:FriMar30202508CEST2012}), the elements of interest for the spectrum read,
\begin{subequations}
  \label{eq:WedMar21033037CET2012}
  \begin{align}
    &\mathbf{w}[01][0,0]=\frac{-1}{M+(i\omega_1-\frac{\Gamma_1}{2})\mathbf{1}}(i\varepsilon_1T_-)\mathbf{v}^\mathrm{ss}\,,\\
    &\mathbf{w}[10][0,0]=\frac{-1}{M+(-i\omega_1-\frac{\Gamma_1}{2})\mathbf{1}}(-i\varepsilon_1T_+)\mathbf{v}^\mathrm{ss}\,.
  \end{align}
\end{subequations}

The final expression is,
\begin{equation}
  \label{eq:WedMar21032339CET2012final}
  \mean{n_1}=\varepsilon_1^2\,\Tr_\mathrm{sys}\Big(\frac{1}{M+(-\Gamma_1)\mathbf{1}}\Big[
  T_+\frac{1}{M+(i\omega_1-\frac{\Gamma_1}{2})\mathbf{1}}T_-+T_-\frac{1}{M+(-i\omega_1-\frac{\Gamma_1}{2})\mathbf{1}}T_+ \Big]\mathbf{v}^\mathrm{ss}\Big)\,.
\end{equation}

\subsection{Two-photon spectrum of emission (two sensors)}

The physical two-photon spectrum in the steady state and at $\tau=0$,
is given by intensity-intensity cross correlations between two sensors
as
\begin{equation}
  \label{eq:FriJan18161900CET2013}
  \mean{n_1n_2}=\mean{\varsigma^\dagger_1\varsigma_1\varsigma^\dagger_2\varsigma_2}=\Tr_\mathrm{sys}(\mathbf{w}[11][11])=\frac{\varepsilon_1^2\varepsilon_2^2}{\Gamma_1 \Gamma_2}(2\pi)^2
  S_{\Gamma_1 \Gamma_2}^{(2)}(\omega_1;\omega_2)\,,
\end{equation}
with,
\begin{equation}
  \mathbf{w}[11][11] =\frac{-1}{M+(-\Gamma_1-\Gamma_2)\mathbf{1}}\Big\{(-i\varepsilon_2 T_+) \mathbf{w}[11][01]
  +(i\varepsilon_2 T_-) \mathbf{w}[11][10]+\left[ 1\leftrightarrow
    2\right]\Big\}\,.
\end{equation}
This solution relies on $\mathbf{w}[11][01]$ and $\mathbf{w}[11][10]$,
which can be expressed in terms of four lower order correlators:
\begin{equation}
  \mathbf{w}[11][01]= \frac{-1}{M+(i\omega_2-\Gamma_1-\frac{\Gamma_2}2)\mathbf{1}} \Big\{i\varepsilon_2 T_- \mathbf{w}[11][00]+i\varepsilon_1
  T_- \mathbf{w}[10][01]-i\varepsilon_1 T_+ \mathbf{w}[01][01]
  \Big\}\,,
\end{equation}
and 
\begin{equation}
  \mathbf{w}[11][10]= \frac{-1}{M+(-i\omega_2-\Gamma_1-\frac{\Gamma_2}2)\mathbf{1}} \Big\{-i\varepsilon_2 T_+ \mathbf{w}[11][00]+i\varepsilon_1
  T_- \mathbf{w}[10][10]-i\varepsilon_1 T_+ \mathbf{w}[01][10]
  \Big\}\,.
\end{equation}
Their solutions are Eq.~(\ref{eq:WedMar21032339CET2012}) and
\begin{equation}
  \mathbf{w}[10][01]= \frac{-1}{M+(-i\omega_1+i\omega_2-\frac{\Gamma_1+\Gamma_2}2)\mathbf{1}} \Big\{i\varepsilon_2 T_- \mathbf{w}[10][00]-i\varepsilon_1 T_+ \mathbf{w}[00][01] \Big\}\,,
\end{equation}
and
\begin{equation}
  \label{eq:ThuMar29195317CEST2012}
  \mathbf{w}[01][01]= \frac{-1}{M+(i\omega_1+i\omega_2-\frac{\Gamma_1+\Gamma_2}2)\mathbf{1}}\Big\{i\varepsilon_1 T_- \mathbf{w}[00][01]-i\varepsilon_2 T_- \mathbf{w}[01][00] \Big\}\,.
\end{equation}
and
\begin{equation}
  \label{eq:FriJan18164304CET2013}
  \mathbf{w}[10][10]= \frac{-1}{M+(-i\omega_1-i\omega_2-\frac{\Gamma_1+\Gamma_2}2)\mathbf{1}} \Big\{-i\varepsilon_1 T_+ \mathbf{w}[00][10]-i\varepsilon_2 T_+ \mathbf{w}[10][00] \Big\}\,.
\end{equation}
By recurrence, we can build the final solution in terms of the system
master equation, $M$, and the sensor parameters directly,
\begin{subequations}
  \label{eq:WedMar21195020CET2012}
  \begin{align}
    \mean{n_1n_2}=&\epsilon_1^2\epsilon_2^2\,\Tr_\mathrm{sys}\Big( \frac{1}{M+(-\Gamma_1-\Gamma_2)\mathbf{1}}\nonumber\\
    \times\Big\{ T_+
    \frac{1}{M+(i\omega_2-\Gamma_1-\frac{\Gamma_2}2)\mathbf{1}}\Big[ &
    T_- \frac{1}{M-\Gamma_1\mathbf{1}}\Big(T_+
    \frac{1}{M+(i\omega_1-\frac{\Gamma_1}{2})\mathbf{1}}T_-+T_- \frac{1}{M+(-i\omega_1-\frac{\Gamma_1}{2})\mathbf{1}}T_+\Big) \nonumber\\
    +& T_-
    \frac{1}{M+(-i\omega_1+i\omega_2-\frac{\Gamma_1+\Gamma_2}2)\mathbf{1}}\Big(
    T_- \frac{1}{M+(-i\omega_1-\frac{\Gamma_1}{2})\mathbf{1}}T_++ T_+ \frac{1}{M+(i\omega_2-\frac{\Gamma_2}{2})\mathbf{1}}T_- \Big) \nonumber\\
    +& T_+
    \frac{1}{M+(i\omega_1+i\omega_2-\frac{\Gamma_1+\Gamma_2}2)\mathbf{1}}\Big(
    T_-\frac{1}{M+(i\omega_2-\frac{\Gamma_2}{2})\mathbf{1}}T_-+ T_- \frac{1}{M+(i\omega_1-\frac{\Gamma_1}{2})\mathbf{1}}T_- \Big) \Big]\nonumber \\
    + T_-
    \frac{1}{M+(-i\omega_2-\Gamma_1-\frac{\Gamma_2}2)\mathbf{1}}\Big[
    & T_+ \frac{1}{M-\Gamma_1\mathbf{1}}\Big(T_+
    \frac{1}{M+(i\omega_1-\frac{\Gamma_1}{2})\mathbf{1}}T_- +T_- \frac{1}{M+(-i\omega_1-\frac{\Gamma_1}{2})\mathbf{1}}T_+\Big) \nonumber\\
    +& T_-  \frac{1}{M+(-i\omega_1-i\omega_2-\frac{\Gamma_1+\Gamma_2}2)\mathbf{1}}
    \Big( T_+ \frac{1}{M+(-i\omega_2-\frac{\Gamma_2}{2})\mathbf{1}}T_++ T_+\frac{1}{M+(-i\omega_1-\frac{\Gamma_1}{2})\mathbf{1}}T_+ \Big) \nonumber\\
    +& T_+
    \frac{1}{M+(i\omega_1-i\omega_2-\frac{\Gamma_1+\Gamma_2}2)\mathbf{1}}
    \Big(T_- \frac{1}{M+(-i\omega_2-\frac{\Gamma_2}{2})\mathbf{1}}T_++ T_+ \frac{1}{M+(i\omega_1-\frac{\Gamma_1}{2})\mathbf{1}}T_-\Big) \Big]
    \nonumber\\
    +[1\leftrightarrow 2]\Big\}\mathbf{v}^\mathrm{ss}\Big)\,.&\tag{A15}
  \end{align}
\end{subequations}
\end{widetext}


\end{document}